# Dispersion and Scaling Law of Dynamic Hysteresis Based on the Landau-Lifshitz-Gilbert Model

Siying Liu, Hongyi Zhang, Hao Yu[*]
Department of Mathematical Sciences, Xi'an Jiaotong-Liverpool University, Suzhou, China

[*]To whom correspondence may be addressed.

## Abstract

Hysteresis dispersion under a varying external field $H_{ex}$ is investigated through numerical simulations based on the Landau-Lifshitz-Gilbert (LLG) equation, indicating the energy dissipation can be determined by $W(\eta) = A\,(f, H_0)$. A linear relation between area of hysteresis and magnitude of external field is discovered. Evolution of hysteresis is also investigated under oscillating external field.

## Introduction

Magnetic phenomenon originates from the magnetic moment of spin and orbit of electron. Magnetic moment in a solid, described as system average ordering parameter $M$ (magnetization) [1, 2], can be aligned to external field and in a processional motion revolving about the axis of field, then hysteresis occurs. Hysteresis is a nonlinear response of a magnet to external field, namely the magnetization of magnet changing with respect to an oscillating magnetic field. A hysteresis loop shows important characteristics of materials, such as saturation magnetization and coercivity, and its area represents the energy dispersion during one revolution. When the applied field is low frequency (quasi-static) and the materials is under equilibrium state, the shape of hysteresis loop can tell us rich information like if the material is soft or hard magnetic materials. Moreover, in some other case, it is more interesting to study hysteresis behavior under fast oscillating field, i.e. dynamic hysteresis, which corresponds to applications in information industry like a high speed magnetic storage, and high frequency magnetic devices.

When a spin system, normally a magnet, is placed in a dynamically varying external field, for instance the sinusoidal $H_{ex} = H_0 Sin(2\pi f t)$, the shape of hysteresis will be influenced by both $H_0$, the amplitude $of\ H_{ex}$ and $f$ the frequency [1,3]. In the extremely high frequency external field, the magnetic system tends to be asymmetrical to origin [1, 2, 4], which tells the importance to investigate the dynamic frequency response of magnetization. There are several approaches to theoretically study the dynamic hysteresis: micromagnetic model based on statistical physics lattice models: Ising or Heisenburg Hamiltonian, or dynamic differential equations. Previously, the hysteresis dispersion has been simulated by $(\psi^2)^3$ model [1, 5, 6] and it was found that the energy dispersion $W(\eta) \propto \tau_1 A(f, H_0)$ where $\tau_1$ is the characteristic time of applied field $H_{ex}$, and $A(f, H_0)$ the area of hysteresis. Also observed from the relationship, the area of hysteresis depends on $f$ and $H_0$. However, high and low frequency of $H_{ex}$ should be considered separately at fixed $H_0$ when to determine $A(f) = kf^\beta$ ($k$ is a constant, and $\beta$ has different values regarding high and low $f$,) [7].

To establish the mathematical model for the spinning magnetic system, Landau-Lifshitz-Gilbert equation (LLG) is utilized to calculate the degree of magnetization $M$ by the external field $H_{ex}$. The LLG differential equation can describe the processional motion of magnetization in material. The dynamic external magnetic field in high frequency applied a torque on nanomagnet moment. The LLG equation formulates weak magnetization behavior and interactions inside magnetic materials, providing an expression of such complex magnetic phenomena as magnetic hysteresis and saturation [4].

In the first section, LLG Equation in a dynamic external magnetic field periodically produces hysteresis and the parameters choices out of stable LLG equation are discussed. Then, different external amplitudes and frequencies, linearly increasing within boundary conditions of stability, produce varying hysteresis surface areas. In the third part, the area of hysteresis indicating the energy loss from magnetization is obtained from the simulations in which the area of hysteresis is influenced by the magnitude and frequency of external field. Finally, the relationship between area and external field frequency based on simulation data is determined using a fitting exponential relation to predict the area in terms of the external field.

## LLG Equation and Model

Original LLG equation is:

$$\frac{dM}{dt} = -\gamma M \times H_{eff} - \alpha \gamma M \times \frac{dM}{dt}$$

where $\gamma$, the gyromagnetic ratio, relates the angular moment to magnet momentum, for example the spinning charged particle like electron in $xy$ plane will generate a z-axis magnetic momentum. Effective field $H_{eff}$ adds a torque on $M$ rotating the momentum spinning with the velocity of $M \times H_{eff}$, which with the damping item $M \times dM/dt$ drive $M$ aligning to the direction of $H_{eff}$ [8, 9].

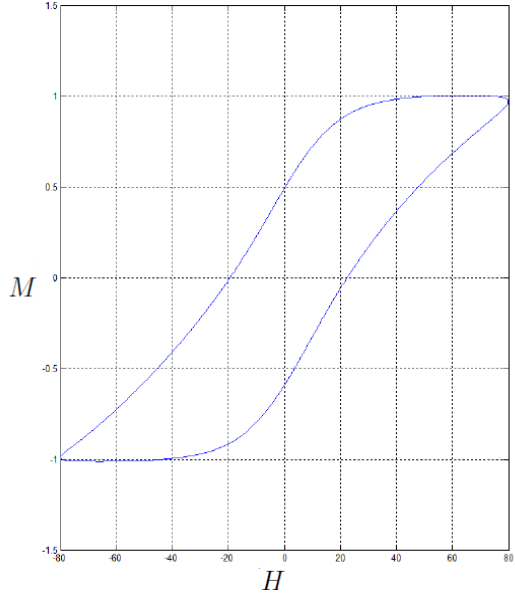

Figure 1: Hysteresis based on $H_0 sin(2\pi ft)$

According to [9], the LLG Equation can be written as:

$$\frac{dM}{dt} = -\gamma M \times H_{eff} - \alpha\gamma M \times (M \times H_{eff})$$

where $\alpha/\gamma \ll 1$, is the damping parameter. Rescale the time and characterize $M$ to a unit vector ($|m| = 1$), and rewrite the LLG equation as

$$\frac{dm}{dt} = -\gamma m \times H_{eff} - \alpha\gamma m \times (m \times H_{eff}).$$

To dynamically investigate the features of LLG Equation, a sinusoidal external field $H_{ex} = H_0 Sin(2\pi ft)$ is applied. The effective field $H_{eff}$ can be determined from the $E(m)$.

$$E(m) = d^2 \int_\Omega |\nabla m|^2 \, dx + Q \int_\Omega \phi(m) dx + \int_{R^3} |H[m]|^2 \, dx - 2 \int_\Omega h_{ex} m \, dx$$

From the energy expression the free energy consists of exchange energy, anisotropies energy, induced field energy, and external field energy [8]. Assume that the material is small soft nanomagnet, $Q = 0$ and $|\Omega| \ll 1$, the effective field vector can be simplified as

$$H_{eff} = d^2 \Delta m + H[m] + h_{ext}$$

To analyse small particles, we apply scaling law to the LLG equation and free energy

equation, we have LLG Equation as

$$m_t = \alpha m \times H_{eff}^{res} - m \times (m \times H_{eff}^{res})$$

and the rescaling effective field as

$$H_{eff}^{res} = \Delta m + \eta^2 H[m] + \eta^2 h_{ext}$$

Based on [2], the effective field vector can be considered as

$$h_{eff} = [h_x, h_y - h_k m_y, h_z - h_{de} m_z]$$

where the external field $h_{ext} = [h_x, h_y, h_z]$, $h_k$ is the anisotropy field in y direction ($h_k = 65Oe$), and $h_{de}$ refers to the demagnetization field in z direction ($h_{de} = 6.5Oe$). Theoretically the two coefficients of LLG equation, $\gamma$ and $\alpha$, could be given randomly, significantly influencing the shape of hysteresis, as shown in Fig.2.

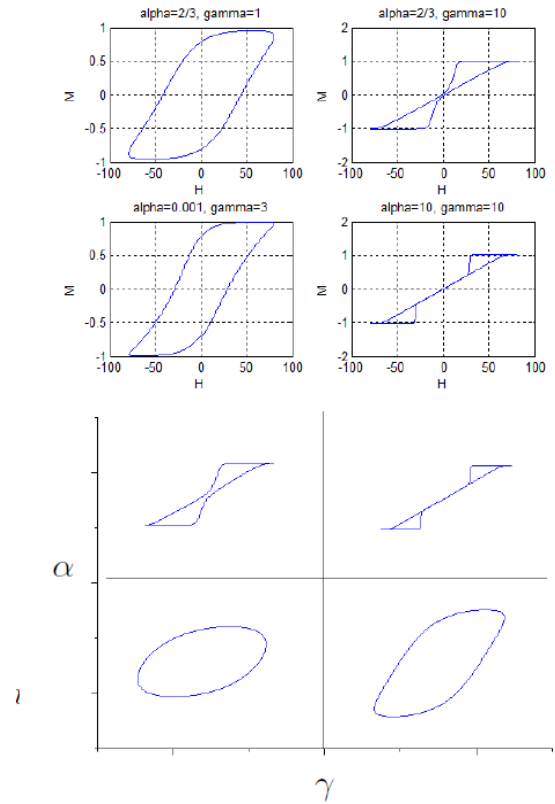

Figure 2: phase transition of hysteresis due to $\gamma$ and $\alpha$.

In regard to the scaling LLG Equation, $\gamma$ decides how fast the spinning angular velocity is while $\alpha$ decides how fast the momentum align to z axis. In the simulation, $\gamma$

=3 is qualified to produce a stable hysteresis. Furthermore $\alpha = 2/3$ ensures that the magnetization reached saturation region ($\gamma$ and $\alpha$ are relative values).

## Hysteresis Dispersion and Scaling

The area of hysteresis is varying when applying external field with different frequencies, reaching the maximum value in a specific frequency. As Fig. 3 shown the hysteresis has different behavior in low frequency region and high frequency region and two separate exponential mathematical models are applied to analyse the frequency response. In regard to the relation between area and external field amplitude, from the linear figure we can deduce that in all frequencies within stability the area is proportional to the field amplitude.

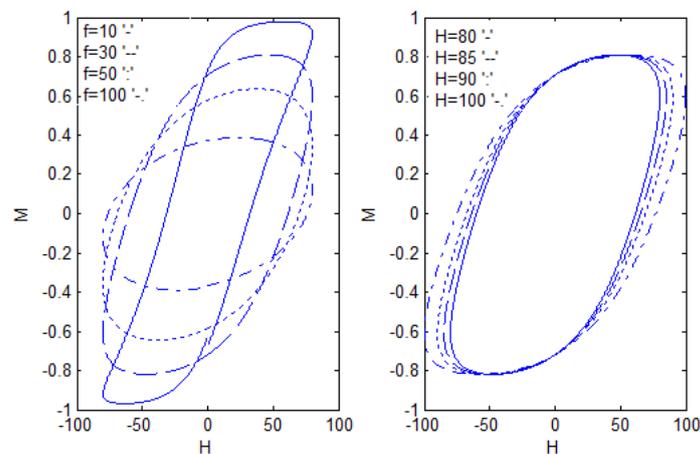

Figure 3. Hysteresis Evolution

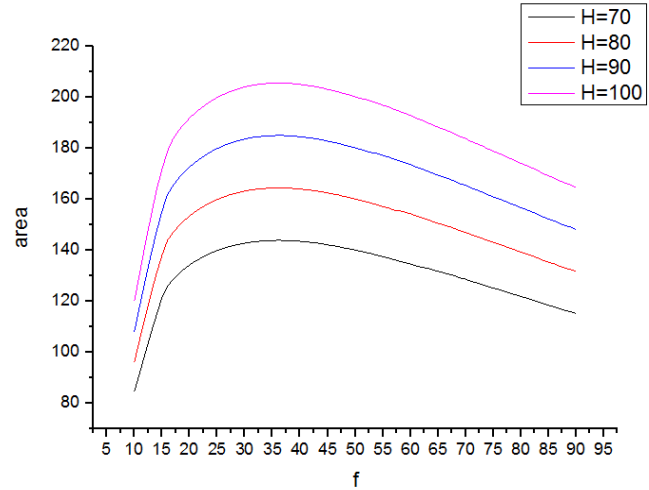

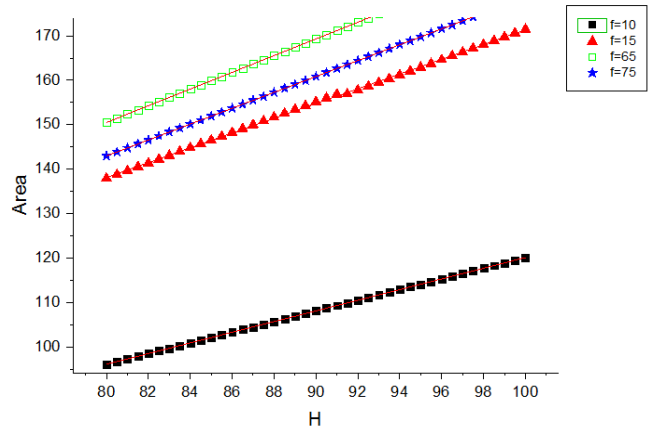

Figure 4. Hysteresis Dispersion

From Fig. 4, at the beginning the area increases as frequency becomes larger until reaching the peak, after which the area decreased as increasing frequency. Area increases along with increasing *H*, showing a linear relation between area and amplitude. With the relationship between area and $H_0$, *f* and according to [2,8,10],

$$A = kH_0^a f^b$$

where *A* is area, *k* is a constant. When find the value of *b*, keep $H_0$ as a constant and to find the value of *a*, *f* stays unchanged. Considering the different behaviors of hysteresis in low frequency and high frequency regions, in this step the curve needs to be divided to two parts, one part is in low frequency region, and another describes the high frequency region. Having the values of *a*, *b* and *k*, the area of hysteresis can be predicted with the given *f* and $H_0$.

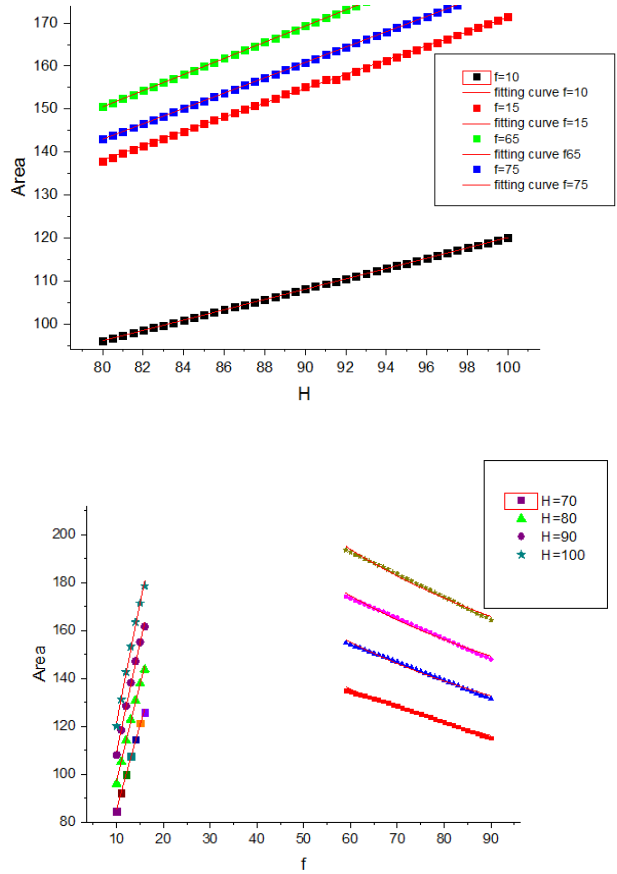

Figure 5: Relations among Area, frequency and amplitude.

The original curve and fitting curve in terms of *f* in the first part is in the upper graph, where $a = 0.85$, then the function is $A = k \times f^{0.85}$. Additionally, the original curve and fitting curve of *f* in the second part is in the lower graph, $a=-0.38$ and, then the function is $A = k \times f^{-0.38}$. Therefore, in the first part of *f*, the final function is $A = 0.17 \times H_0 \times f^{0.85}$, and in the second part of *f*, the final function is $A = 9.19 \times H_0 \times f^{-0.38}$. Finally, the values of the coefficient $\gamma$ and $\alpha$ are changed to check whether the coefficients of function $A = kf^a H_0^b$ may change with the changing of $\gamma$ and $\alpha$. With changing the values of $\gamma$ and $\alpha$, the results are found that the coefficients will be changed with varying $\gamma$ and $\alpha$.

**Conclusion**

This article provides an understanding on Landau-Lifshitz-Gilbert Equation and deduces the energy dissipated every loop in material magnetization. Meanwhile, the numerical simulations generate the hysteresis curves in different conditions involving evolution. Based on current simulation, magnetization is external field dependent and influenced by frequency, especially extremely low/high frequency leading to unstable asymmetric to origin. The resonant frequency corresponding to the maximum area of hysteresis will generate a hysteresis loop with best saturation and good satbility. The energy dissipated under dynamic external field can be predicted by scaling law, which provides an approach to design magnetic materials in the spintronics-related technology.

## Acknowledgement

This work was supported by grants from the National Natural Science Foundation of China (No. 11204245), and the Natural Science Foundation of Jiangsu Province (No. BK2012637).